# A Semi-Structured Tailoring-Driven Approach for ERP Selection

Abdelilah Khaled[1] and Mohammed Abdou Janati Idrissi[2]

[1] TIME Research Team, ENSIAS/UM5 Souissi University, Rabat, Morocco

[2] TIME Research Team, ENSIAS/UM5 Souissi University, Rabat, Morocco

**Abstract**
It has been widely reported that selecting an inappropriate system is a major reason for ERP implementation failures. The selection of an ERP system is therefore critical. While the number of papers related to ERP implementation is substantial, ERP evaluation and selection approaches have received few attention. Motivated by the adaptation concept of the ERP systems, we propose in this paper a semi-structured approach for ERP system selection that differs from existing models in that it has a more holistic focus by simultaneously 1) considering the anticipated fitness of ERP solutions after the optimal resolution, within limited resources, of a set of the identified mismatches and 2) evaluating candidate products according to both functional and non-functional requirements. The approach consists of an iterative selection process model and an evaluation methodology based on 0-1 linear programming and MACBETH technique to elaborate multi-criteria performance expressions.

## 1. Introduction

Enterprise Resource Planning (ERP) system is a configurable and customizable software solution that provides integrated information processing which spans across both functional and technical departments in the organization. It contains a set of standard modules that support different functional areas and consolidate most business operations into a uniform system environment [1]. According to [2], in order to gain competitive edge or to sustain their market share in a highly severe market competition, many organizations around the world have shifted from in house development of business systems to the purchase of ready to use ERP solutions. The reason behind this is the ability of this kind of software to embed the world wide best practices adopted in different management areas, streamline business processes, cut costs, improve quality, produce real time management information, and increase flexibility.

Over the years, despite the desirable benefits of such solutions, many organizations have faced serious implementation failures ([3, 4]). To tackle this issue, many researchers and practitioners have attempted to address this issue by focusing on problems related to both implementation and post-implementation stages of the acquisition process, but to a lesser extent on those related to the pre-implementation one, ignoring the way selection decisions are taken and their appropriateness regarding the acquisition of ERP systems.

Nevertheless, the level of success rate remains low despite a rich literature devoted to implementation models. It seems that research in the first two directions brings only a partial solution to the problem. Needless to say that it must be supplemented by the development of new evaluation and selection approaches that help identifying the most promising solution that best meets the organizations' requirements, prior to its acquisition.

Actually, because of the complexity and diversity of ERP products available on the market, choosing this solution is a non-obvious task. The rich features provided by each product and the sophistication of the organization's requirements have made it almost impossible to choose a suitable solution without using systematic selection approaches.

We argue that in the ERP selection process, two important key factors must be considered. Firstly, the selection decision should be based on both functional and non functional characteristics of the candidate solutions. In fact, the chosen solution is often a trade-off between the satisfactions of functional requirements and other aspects that include technical performance, total cost of ownership, and vendor's characteristics of each alternative. Secondly, ERP tailoring is an important feature that must be considered during the selection process. Unlike traditional software systems, ERP solutions are customizable, which gives them more flexibility to best meet the requirements of each organizational context. Hence, ERP evaluation must undoubtedly considers the maximum anticipated fitness of each alternative after an optimal resolution, within a limited budget, of a subset of the identified mismatches.

Despite their importance, the selection approaches proposed in the literature, still fail to deal simultaneously with the two aforementioned factors. In this regard, this paper intends to develop a semi-structured tailoring-driven approach for ERP selection based on functional and non-functional requirements. It is based on both a 0-1 integer programming model and MACBETH cardinal scales in order to construct performance expressions related to ERP alternatives. The remainder of this paper is structured as follows: we start by giving a short overview of the literature related to ERP selection. In the second section, we introduce our proposed approach and its underlying selection process. Finally, the last section provides conclusions and suggestions for future research.

## 2. Literature Review

During the last decade, there has been a developing body of academic literature that addresses software selection issues in general. Some of the proposed approaches are restricted only to ERP software ([5, 6, 7]), while others are geared toward a wide range of software systems called COTS (Commercial Off The Shelf) solutions ([8, 9]) and could be applied as well to the ERP case. Even though, there is no agreed upon approach for ERP selection in the literature, all the existent approaches have some key steps in common that might be iterative and overlapping ([10]):

**Step 1:** Determine selection criteria based on the organization's requirements.
**Step 2:** Search for the most promising ERP solutions and perform preliminary screening based on must have criteria.
**Step 3:** Evaluate candidate products according to the evaluation criteria.
**Step 4:** Choose the best-fit ERP product.

The author of reference [11] suggests that the main problems behind software selection failure are 1) not well considering organization's requirements and 2) not using appropriate decision models in the decision process. Actually, organization's requirements could be classified into two categories: functional requirements that capture the intended business functionalities that the ERP system is required to support, and non functional ones that describe features differentiating the solution from the other alternatives ([12]).

Current methods for ERP selection fail to effectively support simultaneously both functional and non functional requirements in the evaluation of such systems. Besides, the existent selection models generally neglect the main adaptation feature of ERP systems in the evaluation of this kind of software. In fact, adaptation aims to tailor the ERP system to support additional functionalities in order to increase fitness with the organization's requirements. Actually, handling mismatches among products features and requirement specifications has a direct impact on enhancing the anticipated functional coverage of the ERP solutions. In this regard, ERP selection decision making should consider the anticipated fitness of the package balanced against its non functional capabilities.

In an attempt to address these shortcomings, we develop, in the next section, our proposed selection approach that addresses the highlighted issues.

## 3. Proposed Approach for ERP Selection

ERP selection is a systematic and repeatable process that aims to identify the most promising solution among those available. According to [13], the final purchase decision is influenced by numerous factors. They include decision-makers' preferences and priorities, anticipated benefits, incurred costs, implementation risk and required time for completion and training. In the remainder of this section, we propose a selection approach that addresses the shortcomings presented in the literature review. An iterative process for requirements acquisition and product evaluation is adopted as it is depicted in (Fig. 1).

3.1 Requirements Identification

It is an important stage on which relies the selection project success. It aims to elicit functional and non functional requirements from various stakeholders. Based on the identified requirements the selection criteria are defined. Functional requirements are defined according to each functional area in the organization. Even though the requirements acquisition should be as well documented as possible, it shouldn't, however, focus on basic functionalities common to all products. Indeed, ERP systems have become mature enough that they support them well. The organization should rather focus on important features that would make difference among the proposed solutions. This reduces both the time and the effort required for the evaluation of the various candidate products. Many requirements elicitations techniques are proposed in the literature ([14]). However we recommend in our approach to use the "use case" formalism of UML (Unified Modeling Language) to describe the functional requirements. The reason behind this is that requirements identification should be performed at a high abstraction level in order to not exclude all possible alternatives. Actually, the more requirements definition focuses on describing functionalities at a low abstraction level, the less likely to find a solution that meets all these specifications.

On the other hand, non-functional requirements are qualities or characteristics of the system that the stakeholders care about, and hence will affect their level of satisfaction.

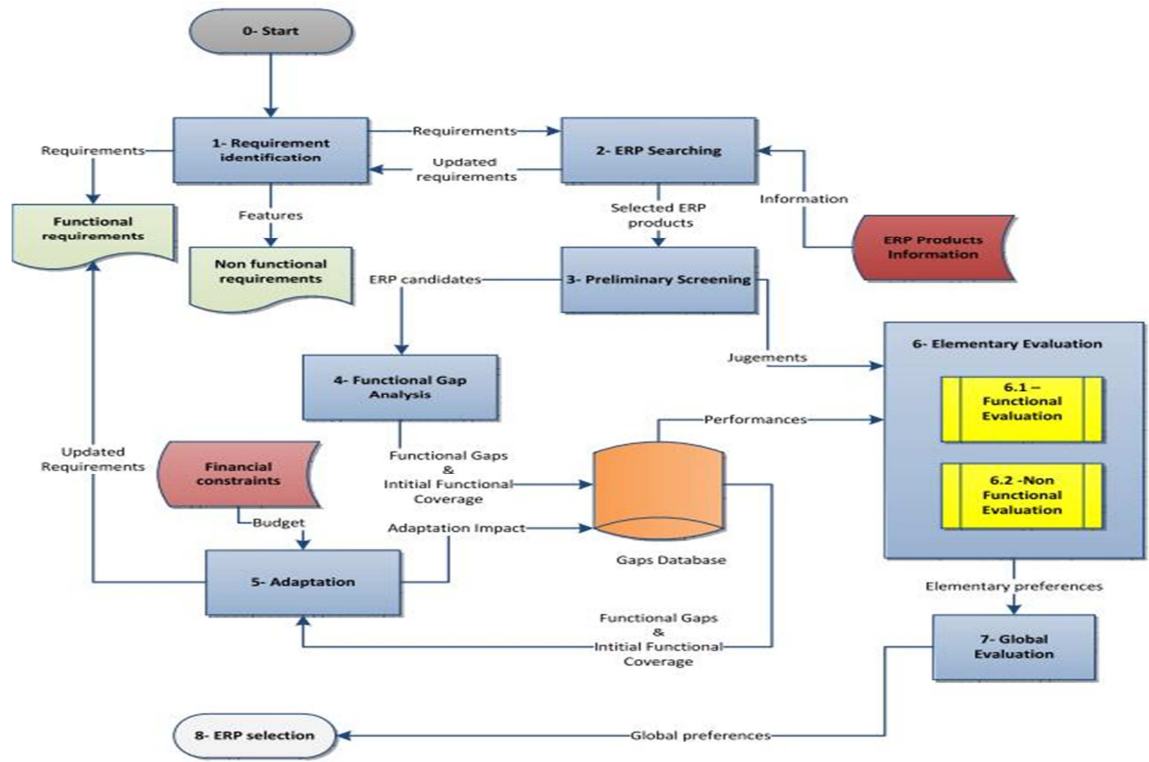

Fig1. Proposed evaluation and selection process

They include financial constraints, technical capabilities like stability and portability and vendor's characteristics. These requirements are translated directly to non functional evaluation criteria.

### 3.2 ERP Searching

In this stage, the organization has to conduct a market research initiative looking for potential ERP systems and to obtain enough information on each expected solution. Many information sources could be used to achieve this objective including internet, benchmarking studies of consulting firms, white papers published by some ERP vendors, functional and technical specifications of ERP products, key conferences on ERP systems and even request of proposals (RFP).

### 3.3 Preliminary Screening

The identified potential candidates are then thoroughly examined in order to narrow the initial products list by choosing the most promising ones for the remainder of the selection process. This reduces both time and effort required for the ERP comparisons. Screening criteria are the minimum requirements expected from candidate products. The author of reference [15] proposes three major screening criteria for this stage: industry type supported by the solution, organization's size to which the solution is destined and technical platform required to support it. Besides, the total cost of ownership and some ERP vendor characteristics should also be considered to screen solutions.

### 3.4 Functional Gap Analysis

This stage aims to identify the mismatches among the functionalities initially supported by each ERP solution and the ones required by the acquiring organization. The authors of reference [16] have defined four matching patterns to describe these mismatches as it is described in table (1).

Besides detecting mismatches among package functionalities and the ones required by the organization, this stage allows the decision maker to detect the advantages of each solution and the functionalities that couldn't be satisfied by any of the alternatives. This gives them a chance to update the initial requirements list and to focus more on the most critical features that are necessary to conduct the organization business. The requirements update process is illustrated in (Fig.3).

Table 1: Matching patterns

| Matching pattern | Description |
|---|---|
| Fulfill | The required functionality is fully satisfied by the package at the target level. |
| Differ | The required functionality is partially satisfied by the package. This pattern occurs when the satisfaction level is acceptable but not optimal. |
| Fail | The satisfaction degree of the required functionality is below the worst acceptable level. |
| Extend | This pattern occurs when the package provides functionalities that are not requested by the organization. It gives rise to the following three impact situations: Neutral, helpful or hurtful. |

In addition, if we denote by $F = \{f_1, f_2, \cdots f_n\}$ the required functionalities set. The decision maker has to construct a satisfaction function denoted by $SAT_i$, as shown in Eq. (1), related to each candidate package $ERP_i$ that determines how much the $ERP_i$ satisfies each functionality $f_j$.

$$SAT_i : F \to [0,1] \quad (1)$$

The determination of these functions will set the stage for the next step of our selection process.

### 3.5 Adaptation

Given the ERP systems adaptability feature, we suggest to evaluate the candidate products based on their fitness after their adaptation to fit the organization requirements. The advantage of doing so is the possibility to consider the best adaptation scenario that could improve the functional coverage of each candidate product within a limited budget. The authors of reference [17] have identified nine tailoring types that we present in table (2). The tailoring types are ranked according to their implementation risk.

Hence, for each functionality $f_j$, a set of adaptation strategies $S_{ijk}, k \in \mathbb{N}$ are to be identified for each $ERP_i$ in order to improve its fitness to the organization requirement. Each adaptation strategy has an implementation risk $r_{ijk}$ and incurs an additional adaptation cost $c_{ijk}$. It aims to improve the satisfaction level related to $f_j$ and $ERP_i$ from $a_{ij}$ to $b_{ijk}$. The parameters of our adaptation model are presented in table (3). The concept of decision binary variables $x_{ijk}$ is inspired by MIHOS ([18]) model of mismatches resolutions. Based on which we propose, for each $ERP_i$ an objective function as it is illustrated in Eq. (2):

$$O_i = \sum_{j \mid a_{ij} \neq 1} w_j (b_{ijk} - a_{ij})(1 - r_{ijk}) x_{ijk} \quad (2)$$

The objective function makes a tradeoff between the fitness enhancement of each functionality defined as $w_j(b_{ijk} - a_{ij})$ when $S_{ijk}$ is chosen and the implementation risk expressed as $(1 - r_{ijk})$. $O_i$ represents how the chosen adaptation strategies improves the fitness of $ERP_i$. Thus, we use the linear optimization model of Eq.(3) in order to determine the values of the binary decision variables $x_{ijk}$.

$$(\forall i) \begin{cases} \max(O_i) \\ (\forall j \mid a_{ij} \neq 1) \sum_k x_{ijk} \leq 1 \\ (\forall j, k \mid a_{ij} \neq 1) \sum_{j,k} x_{ijk} c_{ijk} \leq c_i \end{cases} \quad (3)$$

$\sum_k x_{ijk} \leq 1$ expresses that only one adaptation strategy must be chosen to resolve an identified mismatch.
$\sum_{j,k} x_{ijk} c_{ijk} \leq c_i$ expresses that the sum of adaptation costs shouldn't exceed the budget $c_i$ allowed to each $ERP_i$.

The optimization model of Eq. (3) is a 0-1 linear programming problem that many existent commercial packages could provide an optimal solution.

Table 2: Tailoring types (Source: [17])

| Tailoring type | Description |
|---|---|
| Configuration (customization, in SAP parlance) | Setting of parameters (or tables), in order to choose between different executions of processes and functions in the software package |
| Bolt-ons | Implementation of third-party package designed to work with ERP system and provide industry-specific functionality |
| Screen masks | Creating of new screen masks for input and output (soft copy) of data |
| Extended reporting | Programming of extended data output and reporting options |
| User exits | Programming of additional software code in an open interface |
| ERP Programming | Programming of additional applications, without changing the source code (using the computer language of the vendor) |
| Interface development | Programming of interfaces to legacy systems or 3rd party products |
| Package code modification | Changing the source-codes ranging from small change to change whole modules |

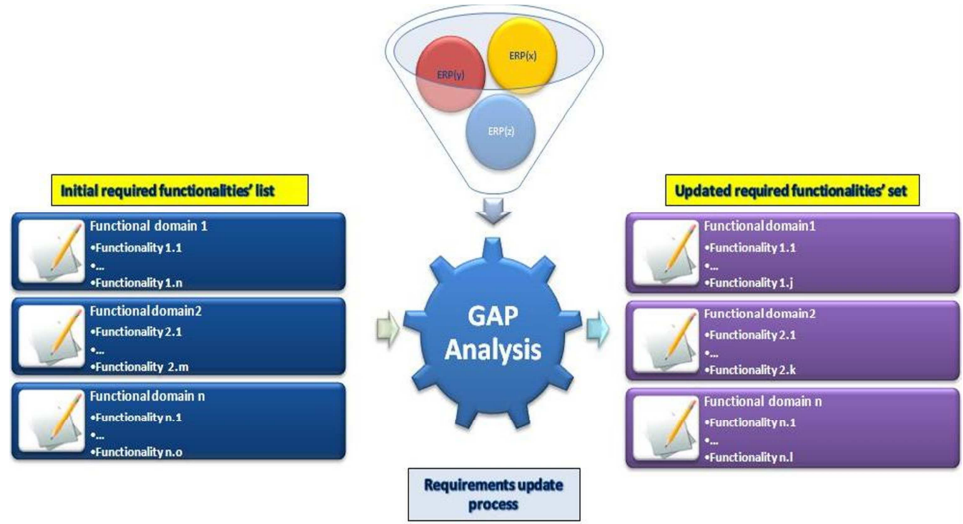

Fig 2. Requirements update process

Table 3: Parameters of the adaptation model

| Parameter | Variable |
|---|---|
| Candidate products | $ERP_i, i = 1 \ldots I$ |
| Required functionalities | $f_j, j = 1 \ldots J$ |
| $f_j$'s importance weight | $w_j, j = 1 \ldots J$ with $\sum_j w_j = 1$ |
| Adaptation strategies related to $f_j$ and $ERP_i$ | $S_{ijk}, k = 1 \ldots K$ |
| Initial satisfaction level related to $f_j$ and $ERP_i$ | $a_{ij} \in [0,1], i = 1 \ldots I$ and $j = 1 \ldots J$ |
| Anticipated satisfaction level related to $f_j$ and $ERP_i$ after tailoring it through $S_{ijk}$ | $b_{ijk} \in [0,1], i = 1 \ldots I, j = 1 \ldots J, k = 1 \ldots K$ |
| Implementation risk related to $S_{ijk}$ | $r_{ijk} \in [0,1], i = 1 \ldots I, j = 1 \ldots J, k = 1 \ldots K$ |
| Additional adaptation cost related to $S_{ijk}$ | $c_{ijk} \in [0,1], i = 1 \ldots I, j = 1 \ldots J, k = 1 \ldots K$ |
| Total budget limit allowed for $ERP_i$ adaptation | $c_i \in \mathbb{R}$ |
| Decision binary variable that indicates whether the adaptation strategy $S_{ijk}$ is chosen or not. $x_{ijk} = 1$ means that the adaptation strategy is chosen and $x_{ijk} = 0$ otherwise. | $x_{ijk} \in \{0,1\}, i = 1 \ldots I, j = 1 \ldots J, k = 1 \ldots K$ |

## 3.6 Elementary Evaluation

The objective of this stage is to determine the various performance expressions related to each ERP candidate product according to the different functional and non functional criteria. In our approach, two techniques are used in order to define these expressions. The first one consists of elaborating mathematical expressions that directly link each ERP to its elementary performance based on quantitative inputs. On the other hand, the second technique uses pair wise comparisons of the alternatives in order to construct these expressions based on qualitative judgments voiced out by the selection team.

Therefore, for the first technique, we rely on the determination of the values related to $x_{ijk}$ variables to define our performance expressions according to the following criteria : 1) the anticipated functional coverage, 2) the adaptation risk related to adaptation strategies, 3) the adaptation additional cost and 4) the adaption degree that would immediately impact the functional coverage after a version update. Table (4) presents the mathematical formulas related to these performance expressions.

For the second technique, we propose to use utility functions in order to construct the performance expressions related to the adopted criteria. Each utility function expresses on a cardinal scale the relative levels of satisfaction associated to the different ERP candidate products. For example, the utility function related to the security criterion assigns numerical values to each ERP system. Hence, the numerical values represent the decision maker satisfaction levels related these systems. Utility functions are defined through the use of semantic scales such as the ones provided by MACBETH ([19]), AHP ([20]) or UTA ([21]). In our case, we suggest using MACBETH (Measuring Attractiveness by a Categorical Based Evaluation Technique) cardinal scales in order to define our utility functions as proposed by ([22]). According to ([19]), MACBETH "*is an interactive approach that requires only qualitative judgments about differences to help a decision maker or a decision-advising group to quantify the relative attractiveness of options. It employs an initial, interactive, questioning procedure that compares two elements at a time, requesting only a qualitative preference judgment*". MACBETH has been widely discussed in its various aspects and it is based on a sound mathematical foundations.

The MACBETH scales construction's procedure relies firstly on the definition of two fictitious values denoted $0_i$ and $1_i$ for each criterion:

- $0_i$: Is defined as the minimal accepted value with respect to $C_i$. An ERP system having a value less than $0_i$ is automatically discarded;
- $1_i$: Is defined as the best value that an ERP system could have regarding $C_i$, which is naturally more attractive than $0_i$.

Similarly, we denote by $ERP_{bad}$ and $ERP_{good}$ two fictitious ERP solutions respectively having $0_i$ and $1_i$ in each criterion:

- $ERP_{good}(1_1, \dots 1_\theta)$ : denotes an ERP solution that has $1_i$ in each criterion.
- $ERP_{bad}(0_1, \dots 0_\theta)$ : denotes an ERP solution that has $0_i$ in each criterion.

MACBETH is based on a questioning procedure in order to construct the cardinal scales. The questioning procedure is as follows:

Let *A* be the set of the ERP options that the decision maker has to choose from, and $B = A \cup \{ERP_{bad}, ERP_{good}\}$.

For each $x, y \in B$ and for each criterion $C_i$, the decision maker is asked to verbally judge the difference of attractiveness between $x$ and $y$ regarding $C_i$. When judging, the decision maker has to choose one of the following categories:

$A_0$-No difference of attractiveness

$A_1$-Very weak difference of attractiveness

$A_2$-Weak difference of attractiveness

$A_3$-Moderate difference of attractiveness

$A_4$-Strong difference of attractiveness

$A_5$-Very strong difference of attractiveness

$A_6$-Extreme difference of attractiveness

However, if the decision maker is unsure about the difference of attractiveness, they may choose the union of several successive categories among these above. 'I do not know' answer is also acceptable and considered in MACBETH as a positive difference of attractiveness.

Table 4: Performance expressions related to the adaptation model

| Performance expression | Description |
|---|---|
| Functional coverage($ERP_i$) = $\sum_j w_j \max_k(\sum_k b_{ijk}x_{ijk}, a_{ij})$ | It represents the total functional coverage of the identified requirements after adaptation. |
| Adaptation risk($ERP_i$) = $1 - \frac{\sum_{j,k \mid a_{ij} \neq 1} \Gamma_{ijk}x_{ijk}}{\sum_{j,k \mid a_{ij} \neq 1} w_j \Delta_{ijk}x_{ijk}}$ <br> With $\Delta_{ijk} = (b_{ijk} - a_{ij})$ and $\Gamma_{ijk} = w_j \Delta_{ijk}(1 - r_{ijk})$ | It represents the risk average associated the all adaptation strategies. |
| Adaptation cost ($ERP_i$) = $\sum_{j,k \mid a_{ij} \neq 1} c_{ijk}x_{ijk}$ | It represents the sum of elementary costs incurred by adaptation strategies |
| Adaptation degree($ERP_i$) = $\sum_{j,k \mid a_{ij} \neq 1} w_j \Delta_{ijk} x_{ijk}\Omega_{ijk}$ <br> With $\Omega_{ijk} = \begin{cases} 0 \text{ if } S_{ijk} \text{ is a simple configuration} \\ 1 \quad\quad\quad\quad\quad\quad\quad\quad\text{Otherwise} \end{cases}$ | It represents the potential functional coverage that the organization would lose immediately after a version update. |

The numerical scales are then obtained by the means of MACBETH judgment matrix. The consistency of this matrix is verified during the expression of the decision maker's preferences. The M-MACBETH (http://www.m-macbeth.com/) software implements the construction process of MACBETH scales. The numerical scales are then extracted thanks to linear programming with the help of two boundary values on each criterion:

$$SCALE_{MACBETH}(ERP_{good}) = 1$$

$$SCALE_{MACBETH}(ERP_{bad}) = 0$$

An example of judgment matrix and it is related MACBETH scale is illustrated in (Fig. 3)

Fig 3. Judgment matrix related to security criterion

### 3.7 Global Evaluation

The global attractiveness of each ERP option is defined as the aggregation of elementary performance expressions defined on each criterion in the previous stage.

In this paper, we continue using the multi criteria decision aid technique of MACBETH to determine this aggregation function that we denote by $\psi$.

MACBETH uses the weighted sum mean (WSM) in order to aggregate the different values $V_i$ defined on the different criteria. Hence $\psi$ is expressed as in Eq. (4):

$$\psi(V_1, .., V_\theta) = \sum_{i=1}^{\theta} \lambda_i \ V_i \text{ with } \sum_{i=1}^{\theta} \lambda_i = 1 \ (4)$$

If we denote by $N = \{1, .. \theta\}$ a simplified notation of the evaluation criteria set C. We easily notice that the aggregation weights can be defined as:

$$\lambda_i = \psi(1_i, 0_{N\setminus i})$$

With ($1_i, 0_{N\setminus i}$) is a performance vector that has 1 on the $i^{th}$ criteria and 0 otherwise. For example

$$\lambda_1 = (1,0,..0)$$
$$\lambda_2 = (0,1,..0)$$
…

Besides, it is easy to verify that:

$$\psi(ERP_{good}) = \psi(1, ... 1) = \sum_{i=1}^{\theta} \lambda_i = 1$$

And

$$\psi(ERP_{bad}) = \psi(0, ... 0) = 0$$

In this regard, MACBETH uses also the same technique of cardinal scales construction in order to determine the different aggregation weights $\lambda_i$. Actually, by ranking the different vectors ($1_i, 0_{N\setminus i}$) including those related to $ERP_{good}$ and $ERP_{bad}$ in a decreasing order of attractiveness, and by using MACBETH categories $A_0$-$A_6$ to judge the difference of attractiveness between each two binary vectors related to $\lambda_i$, we can construct a MACBETH cardinal scale that reflects the relative importance of each criterion. The determination of the different weights $\lambda_i$ is performed under the condition of Eq.5

$$\sum_{i=1}^{\theta} \lambda_i = 1 \ (5)$$

To illustrate this concept, let's suppose that an organization has to evaluate three ERP solutions: SAP, ORACLE and Microsoft Dynamics, according to four criteria as illustrated in (Fig. 4).

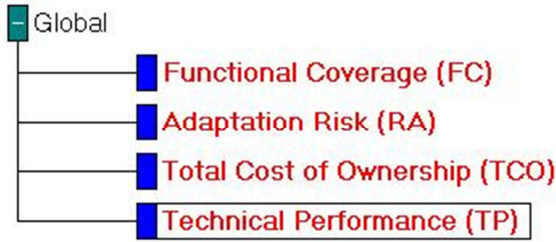

Fig 4. The organization's selection criteria

Then, the decision maker is asked to indicate a decreasing order of preferences regarding ERP options according to each of the four criteria. He has also to judge the difference of attractiveness between each two different ERP options including the two fictitious options: $ERP_{good}$ and $ERP_{bad}$ as it is illustrated in (Fig. 5)

Fig 5. Judgment matrix related to the selection criteria

MACBETH software checks the consistency of the judgments presented by the decision maker and suggests four cardinal scales representing differences of attractiveness among candidate products.

In order to determine the aggregation weights, the decision maker has to judge the differences of attractiveness between the ($1_i$, $0_{N\setminus i}$) vectors denoted by the following reference profiles[FC], [RA], [TP] and [TCO] in (Fig. 6)

Fig 6. Judgment matrix related to the reference profiles

Finally, by using the aggregation function illustrated in Eq. 4, MACBETH determines the global performance related to each candidate product. It provides the final ranking of the ERP options related to their performances and the defined criteria's weights. According to the judgments provided by the decision maker, the ranking of the options shown in the (Fig. 7) suggests that SAP is the best ERP system for this organization.

Fig. 7: Criteria weights and ERP scores

## 4. Conclusion

This paper presented a semi-structured tailoring-driven approach for ERP selection based on the concept of anticipated fitness. A multi-staged evaluation and selection process is then proposed. Besides, the proposed approach is based on both functional and non-functional requirements in order to measure the satisfaction degree of the candidate products. A new evaluation methodology based on 0-1 linear programming and MACBETH technique are used to systematically construct multi-criteria performance expressions related to the different options. Finally, MACBETH aggregation model is then used to rank the initial ERP candidate products 'list. As future perspectives, we propose to apply our approach against real case studies in order to get feedbacks about its validity and improve the evaluation model by reducing the effort required to apply it.

**Abdelilah Khaled** is a software development engineer graduated from ENSIAS (National Higher School for Computer Science and System Analysis) that belongs to the Mohamed V Souissi University. He is also a Ph.D. candidate at the same university. He conducts active research in information technology related to the elaboration of multicriteria decision models that are destined to select and evaluate enterprise systems and COTS (Commercial Off The Shelf) products. Abdelilah Khaled has developed an expertise in software development and business solutions integration. He has worked as a consultant and advisor to various companies and organisations. His works was published in a number of management information systems and computer science academic journals and conference publications.

**Mohammed Abdou Janati Idrissi** holds a PhD degree in Mathematics and Computer Sciences. He is a professor and researcher at ENSIAS (National Higher School for Computer Science and System analysis), Mohamed V Souissi University, Rabat,

Morocco. Prof. Janati is a head of the Decision Aid Department at ENSIAS. His research focuses on decision-aid methods, networks optimization and project management. In addition to his academic work, his engagements include providing expertise to a range of project management, decision aid, and business intelligence projects for many large companies. His works was published in a number of management information systems and computer science academic journals, conference publications, and books.